\documentstyle[sprocl]{article}

\begin{document}

\title{INTEGRABLE SYSTEM \\ 
         AND \\
       $N=2$ SUPERSYMMETRIC YANG-MILLS THEORY}
\footnote{
Talk presented at the Workshop 
``Frontiers in Quantum Field Theory'' 
in honor of the 60th birthday of 
Prof. Keiji Kikkawa, Osaka, Japan, 
December 1995.}

\author{TOSHIO NAKATSU}
\address{Department of Mathematics and Physics,
               Ritsumeikan University, \\
        Kusatsu, Shiga 525-77, Japan}
\author{KANEHISA TAKASAKI}
\address{Department of Fundamental Sciences,
           Faculty of Integrated Human Studies,\\ 
              Kyoto University, \\
         Yoshida-Nihonmatsu-cho, Sakyo-ku, Kyoto 606, Japan }

\maketitle\abstracts{
We study the exact solution of $N=2$ supersymmetric $SU(N)$  Yang-Mills  
theory in the framework of the Whitham-Toda hierarchy. 
We show that it is in fact obtainable by modulating the solution  
of the (generalized) Toda lattice associated with moduli of curves. 
The relation between the holomorphic pre-potential of the low energy 
effective action  and the $\tau$-function of the (generalized) Toda lattice 
is also clarified.}

\section{}

    Recently Seiberg and Witten \cite{SW} obtained exact expressions 
for the metric 
on moduli space and the dyon spectrum of $N=2$ supersymmetric 
$SU(2)$ Yang-Mills theory 
by using a version of the Montonen-Olive duality \cite{duality} and  
holomorphy 
\cite{seiberg} 
of 4$d$ supersymmetric theories. 
Their approach has been generalized to the case of other Lie group 
\cite{N=2SUSY}~~\cite{N=2SUSY2}.  
Especially surprising in these results is 
unexpected emergence of elliptic (or hyperelliptic) curves and their periods. 
Although these objects appear 
in  the course of determining the holomorphic pre-potential $\cal F$ 
of the exact low energy effective actions,
physical siginificance of the curves themselves is unclear yet.  
It will be important to clarify their physical role. 
An interesting step in this direction has been taken  
\cite{integrable} from the view of integrable systems, 
in which the correspondence between the Seiberg-Witten solution \cite{SW} 
and a Gurevich-Pitaevsky solution \cite{GP} 
to the elliptic Whitham equations \cite{Whitham} 
is pointed out.

~

            The exact solution (Seiberg-Witten type) of N=2 supersymmetric 
$SU(N)$ Yang-Mills 
theory is described by the following data 
\cite{SW}~~\cite{N=2SUSY}:
\newtheorem{data}{Data}
\begin{data}
"family of the hyperelliptic curves"
\begin{eqnarray}
\mbox{$\cal C$}: y^2 = P(x)^2 - \Lambda^{2N}~,~~~~~~
 P(x)=x^N+\sum_{k=0}^{N-2}u_{N-k}x^k, 
\label{curve C} 
\end{eqnarray}
where $u=(u_2, \cdots, u_N)$ are the order parameters 
and $\Lambda$ is the $\lambda$-parameter of this theory.
\end{data}
\begin{data}
"the meromorphic differential on $\mbox{$\cal C$}$"
\begin{eqnarray}
dS = \frac{x \frac{dP(x)}{dx}}{y}dx~~.~~~~  
\label{dS-1}
\end{eqnarray}
\end{data}
Using these two data the holomorphic pre-potential 
$\cal F$$=$$\cal F$$(a)$ 
is prescribed from the relation, 
\begin{eqnarray}
\frac{\partial \mbox{$\cal F$}}{\partial a_i}=\oint_{\beta_i}dS,~~~~~~~
a_i=\oint_{\alpha_i}dS~~~~~(1 \leq i \leq N-1) 
\end{eqnarray}
where 
$\alpha_i$,$\beta_i$ are the standard homology 
cycles of the curve $\mbox{$\cal C$}$.

~

            In this talk we discuss  
the exact solution of $N=2$ SUSY $SU(N)$ 
Yang-Mills theory in the framework of the Whitham equations 
\cite{Whitham} \cite{Whitham hierarchy2}  
and clarify the relation with  Toda lattice. 
Especially we would like to explain 
$(i)$
the integrable structure of $SU(N)$ Seiberg-Witten solution 
has its origin in (generalized) $N$-periodic Toda lattice (or chain), 
and 
$(ii)$
the holomorphic pre-potential $\cal F$$(a)$ is obtainable from the 
$\tau$-function of this Toda lattice (or chain) by modulation 
(Whitham's averaging method \cite{Whitham}).

\section{}
\subsection{}

            Let us begin by dscribing the integrable structure which 
already appears implicitly in these data. 
For the convenience we introduce the meromorphic function $h$ by  
\begin{eqnarray}
h=y+P(x) ,
\label{def of h}
\end{eqnarray} 
and consider the partial-derivation of $dS$ by the moduli parameter $a_i$ 
with fixing this function $h$. It satisfies
\begin{eqnarray}
\left.\frac{\partial}{\partial a_i} dS  
~~\right|_{fix~h} = dz_i, 
\label{integrable system}
\end{eqnarray}
where $dz_i$ is the normalized holomorphic differential, 
$^{i.e}$ $\oint_{\alpha_j}dz_i=\delta_{i,j}$. Therefore these 
holomorphic differentials satisfy 
\begin{eqnarray}
\frac{\partial}{\partial a_i} dz_j =
\frac{\partial}{\partial a_j} dz_i~~~~~(1 \leq i,j \leq N-1),   
\label{FFM-1}
\end{eqnarray}
where the derivations by the moduli parameters 
$a=(a_1,\cdots,a_{N-1})$ are performed by fixing 
$h$. These equations are the compatibility conditions for the system 
of holomorphic differentials under their evolutions 
by the moduli parameters 
$a$. So these compatibility conditions define a 
integrable system. 
$dS$ gives a solution for this system.

\subsection{}

      Nextly let us explain the moduli curve 
{\em $\cal C$}
in the data can be understood 
\cite{M-W}~ \cite{N-T} as the spectral curve 
\cite{Toda curve} of ($N$-periodic) Toda chain. 
($N$-periodic) Toda chain is a 1$d$ integrable system 
defined by the equations,
\begin{eqnarray}
\partial_t^2\phi_n=
e^{-(\phi_{n+1}-\phi_{n})}
-e^{-(\phi_{n}-\phi_{n-1})},~~~~
\phi_{n+N}=\phi_n~~~~~(n \in Z).
\label{Toda eq.1}
\end{eqnarray}
The Lax representation of ($N$-periodic) Toda chain is given by 
\begin{eqnarray}
\partial_t \vec{\Psi}=B \vec{\Psi}~~~,~~~
C \vec{\Psi}=x \vec{\Psi}~~~,
\label{Lax}
\end{eqnarray}
where $\vec{\Psi}=^t(\Psi(1), \cdots, \Psi(N))$. 
$B=(B_{i~j})$ and $C=(C_{i~j})$ are $N$x$N$ matrices of 
the following forms, 
\begin{eqnarray}
B_{i~j}&=&d_i\delta_{i+1~j}-d_j\delta_{i~j+1}
        +hd_N\delta_{i~N}\delta_{j~1}
                 -\frac{d_N}{h}\delta_{i~1}\delta_{j~N},
\nonumber \\
C_{i~j}&=&d_i\delta_{i+1~j}+d_j\delta_{i~j+1}
             +b_i\delta_{i~j}
        +hd_N\delta_{i~N}\delta_{j~1}
          +\frac{d_N}{h}\delta_{i~1}\delta_{j~N} . 
\label{BC}
\end{eqnarray}   
$x$ and $h$ in (\ref{Lax}) and (\ref{BC}) are 
the scalar variables independent of $t$. 
By identifying $b_n$ and $d_n$
with $\partial_t \phi_n/2$ and 
$exp\left\{-(\phi_{n+1}-\phi_n)/2\right\}/2$ 
respectively,   
the compatibility condition of equations (\ref{Lax}),   
$\partial_tC=$$\mbox{$[$}$$B,C$$\mbox{$]$}$,  
gives Toda chain equation (\ref{Toda eq.1}).

          The spectral curve of linear system 
(\ref{Toda eq.1}) can be determined from the 
condition ,
\begin{eqnarray}
det \left|_{~} x-C \right|_{~}=0,
\end{eqnarray}
which we can solve with respect to $h$ :
\begin{eqnarray}
h=\frac{\pm y+P(x)}{\Lambda^N},
\label{Toda h}
\end{eqnarray}
where 
$P(x)$ is the monic polynomial of degree $N$ 
and $\Lambda^N=2d_1\cdots d_N$.
$y$ in (\ref{Toda h}) is given by 
\begin{eqnarray}
y^2=P(x)^2-\Lambda^{2N},
\label{Toda spec}
\end{eqnarray}
which defines the spectral curve. 
It coincides with the moduli curve {\em $\cal C$} 
(\ref{curve C}).

\subsection{}

           The solution of linear system (\ref{Lax}), which becomes a 
meromorphic function on the spectral curve (except at the divisor of $h$), 
has been constructed by Krichever \cite{Toda curve}. 
It is shown \cite{N-T} that there 
exist $N-1$ additional charges and that the solution $\Psi(n)$ is allowed 
to depend on the corresponding new parameters 
$\theta=(\theta_1,\cdots,\theta_{N-1})$. It has the following form,
\begin{eqnarray}
\Psi(n,t,\theta)(p)
&=&exp \{-n\int^pdB_0+t\int^pdB_1
       +\sqrt{-1}\sum_{i=1}^{N-1}\theta_i\int^pdz_i\}  \nonumber \\ 
 &&~~~~~~~~~~~~~~~~~~\times \mbox{theta functions}, 
\label{wave function}
\end{eqnarray}
where $dB_{0,1}$ are the normalized meromorphic differentials specified 
by their behaviors around the divisor of $h$.

           Now we will describe the modulation of 
wave function (\ref{wave function}). Suppose there exists an observer whose 
length scales $T_0,T_1$ and $a_i$($1 \leq i \leq N-1$) are very slow relative 
to $n,t$ and $\theta_i$ ; 
\begin{eqnarray}
T_0=-\epsilon n,~~~T_1=\epsilon t,~~~a_i=\sqrt{-1}\epsilon \theta_i, 
~~~~~(\epsilon \ll 1)
\label{slow variables}
\end{eqnarray}
and then the spectral curve is "slowly" varying on these observer's scales. 
On this circumstance the wave function will have the following WKB form,
\begin{eqnarray}
\Psi(n,t,\theta)=A(T_0,T_1,a,\epsilon)e^{\frac{1}{\epsilon}S(T_0,T_1,a)}.
\end{eqnarray}
Taking up the 0-th order approximation leads \cite{N-T} 
to the existence of a meromorphic differential, $dS$, which satisfies 
\begin{eqnarray}
\frac{\partial}{\partial a_i}dS=dz_i,~~~
\frac{\partial}{\partial T_0}dS=dB_0,~~~
\frac{\partial}{\partial T_1}dS=dB_1.
\label{modulation}
\end{eqnarray} 
By setting $ T_0=0 $ and $ T_1=1 $ 
these equations reduce to those in (\ref{integrable system}).  
The meromorphic differential $dS$ 
in data (\ref{dS-1}) can be obtainable from the modulation of 
wave function (\ref{wave function}).

\subsection{}

             According to the prescription  \cite{Toda lattice hierarchy}
the $\tau$-function can be also introduced by studying the behaviors of 
wave function 
(\ref{wave function}) around  the divisor of $h$. 
Let us consider the modulation of this $\tau$-function.
With the same spirit as for the case of wave function 
we can derive the expansion,
\begin{eqnarray}
\ln \tau(n,t,\theta)=\epsilon^{-2}
 \left( F(T_0,T_1,a)+\mbox{$\cal O$}(\epsilon)\right).
\label{modulation2}
\end{eqnarray}
The quantity, $F$, which is the 0-th order approximation, 
turns out \cite{N-T} to satisfy  
\begin{eqnarray}
 F(T_0=0,T_1=1,a)  =\mbox{$\cal F$}(a),
\end{eqnarray}
where $\cal F$ is the pre-potential of 
$N=2$ SUSY $SU(N)$ Yang Mills theory.

\section*{Acknowledgments}
It was a very honor for us 
to participate in this workshop 
and to celebrate Prof. Kikkawa's 60th 
birthday.

\end{document}